\documentclass[12pt,epsfig]{article}
\usepackage{graphics,natbib,amssymb}
\usepackage{epsfig}
\usepackage{graphics}
\begin{document}
\begin{center}
{\Large  {\bf Tunneling times through barrier with inelasticity}}
\end{center}
\vskip .3cm
\begin{center}
 Arya Paul$^\dagger$, Arnab Saha$^\dagger$, 
 Swarnali Bandopadhyay$^\ast$
and Binayak Dutta-Roy$^\S$
\end{center}
\vskip .5cm
\noindent $\dagger$ S.N.Bose National Centre for Basic Sciences, 
Block-JD, Sector-III, Salt Lake City, Kolkata-700098, India.
\vskip .2cm
\noindent $\ast$ Department of Physics, Ben-Gurion University, Beer-Sheva 84105, Israel.

\vskip .2cm
\noindent $\S$ H19 Baishnab Ghata Patuli (Flat 32), Calcutta 700 094, India.
\vskip .4cm 

\begin{abstract}
Tunneling delay times of wavepackets in quantum mechanical penetration of 
rectangular barriers have long been known to show a perplexing independence 
with respect to the width of the barrier. This also has relevence to the 
transmission of evanescent waves in optics. Some authors have claimed that in 
the presence of absorption or inelastic channels (which they model by taking a 
complex barrier potential) this effect no longer exists, in that the time delay becomes proportional to the barrier width. Taking the point of view that 
complex potentials imply non-Hermitian Hamiltonians and are as such fraught 
with conceptual pit-falls particularly in connection to problems involving 
time evolution, we have constructed a two-channel model which does not suffer 
from such maladies in order to examine this issue.  We find that the 
conclusions arrived at by the earlier authors need to be qualified.
\end{abstract}

\vskip .4cm

\begin{center}
{\bf 1. Introduction}
\end{center}

\vskip .2cm

The question of tunneling time [or, `time spent' by a particle of mass $m$ 
incident with energy($E$) less than the height($V_0$) of a barrier of width 
$L$ in the classically forbidden region ($E<V_0$)] and the analogous situation 
of evanescent waves in optics, have in recent years attracted considerable 
attention even though interest in such matters go back to the forties and 
fifties of the last century as witnessed by Eisenbud's Ph.D dissertation$^1$ 
with E.P.Wigner at Princeton in 1948 and Wigner's own paper$^2$ in 1955. 
David Bohm's book on Quantum Theory$^3$ published in 1951 also throws light on 
such issues.

In the present study we shall focus our efforts on one particular aspect of 
tunneling phenomenon [now-a-days referred to as the Hartman$^4$-Fletcher$^5$ 
effect], namely, the counter intuitive conclusion that for a sufficiently 
opaque barrier tunneling delays are independent of the barrier width. 
By the opaque condition, we imply that the mean energy of the incident wave 
packet is much smaller than the height of the potential barrier. The tunneling 
time ($\tau$) is defined in this context by the time taken by the peak of the 
incident wave-packet to traverse the classically forbidden region and to emerge as the transmitted wave-packet. Consider the evolution of an incident localized wave-packet described by 
$$
\int{G_{k_0}(k)exp(ikx-iEt/\hbar)dk}
$$ 
where $G_{k_0}(k)$ is a normalised function of $k$ (say a Gaussian) peaked 
about the mean momentum $\hbar k_0$. If this were allowed to evolve freely 
the peak would travel with a velocity $\hbar k_0/m$. However with a barrier 
present an incident plane wave 
$$
\frac{1}{\sqrt{2\pi}}e^{ikx-iEt/\hbar}
$$ 
(corresponding to a particle of energy $E=\frac{\hbar^2 k^2}{2m}$, where $k$ is the wave vector) would after transmission through it emerge as
$$
\frac{1}{\sqrt{2\pi}}a_T(k) e^{i(kx-Et)/\hbar}
$$
where $a_T=|a_T|e^{i\delta}$ is the transmission amplitude (with $\delta$  
its associated phase). Accordingly, the transmitted wave-packet would be
$$
\int{G_{k_0}(k)|a_T|e^{ikx-iEt/\hbar+\delta(k)}dk}
$$
Suppose for concreteness, we consider a barrier between $x=0$ and $x=L$, then the time($\tau$) at which the peak of the packet will emerge from the barrier at $x=L$ will be given using the method of stationary phase by $\frac{d}{dk}[kL-E\tau/\hbar+\delta(k)]=0$ or
$$
\tau=\hbar\frac{d\delta}{dE}+\frac{L}{(\hbar k/m)}.
\eqno(1)
$$
All quantities are to be evaluated at the peak momentum $\hbar k_0$ and the 
corresponding energy $\hbar^2 k^2_0/2m$. For notational brevity, we suppress 
the subscript $0$.
In our discussion, this will be taken to be a measure of the tunneling time.

Specialising to a rectangular barrier,
$$
V(x)=V_0   \phantom{xxxxxx}    \textrm{for} \phantom{xxxx} 0\leq x\leq L,
\eqno(2a)
$$
$$
V(x)=0      \phantom{xxxxxx}    \textrm{elsewhere},
\eqno(2b)
$$
we are to first search for the solution of the Schr\"odinger equation 
corresponding to stationary states of energy $E$,
$$
\left[-\frac{\hbar^2}{2m}\frac{d^2}{dx^2}+V(x)\right]\psi(x)=E\psi(x)
\eqno(3)
$$
with the required condition describing reflection and transmission, we have 
on the two sides of the barrier
$$
\psi(x)=e^{ikx}+a_R(k)e^{-ikx} \phantom{xxx}    \textrm{for}\quad x<0  \\
\eqno(4a)
$$
$$
\psi(x)=a_T(k)e^{ikx}    \phantom{xxxxxxxxx}    \textrm{for}\quad x>L ,
\eqno(4b)
$$
where $a_R(k)$ and $a_T(k)$ are the reflection and transmission amplitudes whose square modulii give the respective coefficients. The solution within the barrier is given by,
$$
\psi(x)=Ae^{qx}+Be^{-qx}
\eqno(4c)
$$
with $q^2=\frac{2m}{\hbar^2}(V_0-E)$. We may as usual proceed to determine $A$, $B$, $a_R$ and $a_T$ by matching $\psi(x)$ and $\frac{d\psi(x)}{dx}$ at the boundaries ($x=0$ and $x=L$) to obtain
$$
A=\frac{-(1+\frac{iq}{k})exp(-qL)}{(1+\frac{q^2}{k^2})sinh(qL)}
\eqno(5a)
$$
$$
B=\frac{(1-\frac{iq}{k})exp(qL)}{(1+\frac{q^2}{k^2})sinh(qL)}
\eqno(5b)
$$
and
$$
 a_T =\frac{4i\frac{k}{q}exp(-ikL)}{exp(-qL)(1+i\frac{k}{q})^2-exp(qL)(1-i\frac{k}{q})^2}
\eqno(6)
$$
and accordingly we arrive at the phase-shift $\delta$.
$$
\delta=tan^{-1}\left[\frac{(q^2-k^2)}{2qk}tanh(qL)\right] -kL,
\eqno(7)
$$
and using eq.(1), we have
$$
\tau=\hbar \frac{d}{dE}tan^{-1}\left[\frac{k^2-q^2}{2qk}tanh{qL}\right],
\eqno(8)
$$
where the cancellation (in the expression for the time delay) of the 
contribution arising from the term $-kL$ in $\delta$ physically represents 
the time the wave-packet would have taken to traverse the distance from $x=0$ 
to $x=L$ if the potential were absent. It is also understood that we must 
evaluate the delay at the energy $E_0$. We suppress as mentioned earlier the 
subscript zero for notational convenience. The tunnel time $\tau$ as given by 
Eq.(8) is plotted as a function of the barrier width L for two values of 
$\frac{V_0}{E}$ and shown in Fig.1. As must be the case, $\tau \rightarrow 0$ 
as $L \rightarrow 0$. But what is more significant is that as 
$L  \rightarrow \infty$ note that $tanh(qL) \rightarrow  1$ and tunnel time 
$\tau$ becomes independent of the length $L$ of the barrier. This is the 
counter-intuitive Hartman$^4$ Effect. The saturated or asymptotic value of 
$\tau$ is given by
$$
\tau_{asymp}=\frac{\hbar}{\sqrt{E(V_0-E)}}=2\frac{m}{\hbar k}\frac{1}{q}
\eqno(9),
$$
that is, twice the time taken to traverse the decay distance $\frac{1}{q}$ 
in the barrier region if the particle were moving with its free velocity 
$\frac{\hbar k}{m}$.
Explicitly the expression for $\tau$ is given by
$$
\tau=\frac{g}{1+g^2}\left(\frac{m}{\hbar q^2}-
\frac{4m}{\hbar (q^2-k^2)}-\frac{m}{\hbar k^2}-
\frac{mL}{\hbar q^2sinh(qL)cosh(qL)}\right)
$$
where $g=\frac{(q^2-k^2)tanh(qL)}{2qk}$.\\
\begin{figure}[h]
\begin{center}
\includegraphics[angle=0,width=6cm]{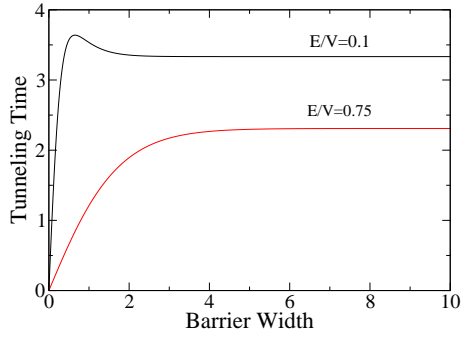}
\caption[Hartman-Fletcher Effect]{The tunnel time ($\tau$) is shown as a 
function of barrier width ($L$) for two values of $E/V_0$ namely $0.1$ 
(dashed curve) and $0.75$ (solid curve).}
\end{center}
\end{figure}

\noindent It is clearly evident both from the graph and 
the expression for $\tau$ that 
for large $L$, $\tau$ is independent of $L$ and hence it attains a saturation 
value. This is what has been known as the Hartman$^4$-Fletcher$^5$ effect for 
decades. What needs to be emphasized is that the tunneling time attains the 
saturation value in two ways depending on the sign of 
$k^2-q^2=\frac{2m}{\hbar^2}(2E-V_0)$. For $k^2-q^2>0$ (i.e. $E>\frac{V_0}{2}$) 
it monotonically increases to reach the saturation value and for $k^2-q^2<0$ 
(i.e. $E<\frac{V_0}{2}$) it reaches the saturation value from above (there is a hump before it attains  saturation). Keeping the next to leading term for large $L$, shows the approach to asymptotia viz. 
$$  
\tau \simeq \tau_{asymp}+8L\left(\frac{1}{\hbar k/m}\right)
\left[\frac{E(V_0-2E)}{V_0^2}\right]e^{-2qL}
\eqno(10)
$$
\begin{figure}[t]
\begin{center}
\includegraphics[angle=0,width=6.0cm]{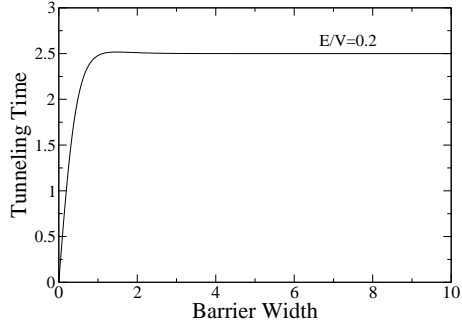}
\caption[]{This graph is for $E/V=0.2$. The error in the position of the maxima of the hump between the actual value and that calculated from Eq(11) is about 
11\%.}
\end{center}
\end{figure}
It is evident from the expression above that there is a exponential approach to the asymptotic limit from below if $E<\frac{V_0}{2}$ ({\it viz.} slope is 
positive with respect to $L$) and from above for $E>\frac{V_0}{2}$.\\
The position of the maximum of the hump ($L_0$) in the latter case is given by
$$
L_0=\frac{1}{4}\left(\frac{\hbar q}{m}\right)\left(\frac{\hbar}{E}\right)
\frac{10E^2-9V_0E+V_0^2}{(2E-V_0)(V_0-E)}
\eqno(11)
$$

\vskip .4cm
\begin{center}
{\bf 2. Hartman Effect in absorbing media}
\end{center}
\vskip .2cm
For investigating the Hartman$^4$-Fletcher$^5$ effect in presence of an 
absorbing media, F.Delgado, Muga \& Rushhaupt$^6$ and 
Fabio Raciti \& Salesi$^7$ phenomenologically introduced an imaginary part 
to the potential in order to describe  absorption. This is in the spirit of the optical model which is often invoked in nuclear physics when there are 
inelastic channels. The relative phase-shift between the transmitted and the 
incident wave-packet in such a situation is obtained by just putting the 
complex potential form in place of $V_0$ in $q$ and inserting it in eq(7). If 
the potential is given by $V=V_0(1-i\lambda)$ within the range $L$ and zero 
elsewhere, then the transmission coefficient ($a_T$) is given by
$$
a_T= \frac{4i\frac{k}{q}e^{ikL}}{e^{-qL}(1+\frac{ik}{q})^2-
e^{qL}(1-\frac{ik}{q})^2}
\eqno(12)
$$
Hence in the large $L$ limit, $a_T$ will take the form of
$$
a_T= \frac{4i\frac{k}{q}e^{ikL}}{e^{qL}(1-\frac{ik}{q})^2}
\eqno(13)
$$
For complex potentials and taking only the first order term in $\lambda$, 
it is seen that
$$
q=q_0\left(1-\frac{i\lambda V_0}{2(V_0-E)}\right)
$$
where $q_0= [\frac{2m}{\hbar^2}(V_0-E)]^{1/2}$.
Hence, we have
$$
\frac{1}{q}\approx\frac{1}{q_0}\left(1+\frac{i\lambda V_0}{2(V_0-E)}\right)
$$
which is readily expressed in a polar form as 
$$
\frac{1}{q}\approx\sqrt{\left(\frac{1}{q_0}\right)^2+
\left(\frac{\lambda V_0}{2q_0(V_0-E)}\right)^2}e^{itan^{-1}
\left(\frac{\lambda V_0}{2(V_0-E)}\right)}
\eqno(14)
$$
Hence $a_T$ in presence of the complex potential will be
$$
a_T=
\frac{-\frac{4ik}{q_0}\left(1+\frac{i\lambda V_0}{2(V_0-E)}\right)e^{-ikL-
q_0\left(1-\frac{i\lambda V_0}{2(V_0-E)}\right)L}}{\left[1-\frac{ik}{q}
\left(1+\frac{i\lambda V_0}{2(V_0-E)}\right)\right]^2}
\eqno(15)
$$
The only $L$ dependence in the phase-shift comes from the numerator of eqn(15) 
and the dependence is of the form of
$$
-kL + \frac{q_0\lambda V_0}{2(V_0-E)}L
$$
Hence the tunneling time $\tau$ for large $L$ is given by
$$
\tau \rightarrow \frac{\lambda V_0}{2(V_0-E)}\frac{k}{q_0}L + \xi
\eqno(16)
$$
where $\xi$ is the $L$-independent factor that comes from the denominator of 
eqn(15). 
It is evident from the expression of $\tau$ that the tunneling time for large 
$L$ is not independent of $L$ and hence we no longer have the Hartman-Fletcher 
effect even for weak absorption (small $\lambda$). Optical experiments by 
Nimtz, Spieker \& Brodowsky$^8$ confirms the absence of Hartman-Fletcher  
effect only for strong absorption and the absence of this effect is not claimed for weak absorption.

Furthermore, one may express doubts regarding the conclusion on tunneling times in the presence of inelastic channels based on the phenomenological device of 
introducing complex potentials (through the paradigm of the optical model)  
particularly when we are dealing with questions regarding time evolution. 
Non-unitary time-evolution using a non-hermitian Hamiltonian could be 
questionable. If randomness or disorder in the system is invoked then one 
must use a density matrix approach to the netire problem. We put forward a 
two-channel formulation for this problem and 
analyze the matter of tunneling time in this format where unitarity is properly maintained.\\

 To obtain a simple realisation of such a situation, consider the elastic 
scattering of a particle by a target system describable in terms of a 
rectangular potential $V_0$ of width $L$. Suppose that the target has an 
excited state of energy $\Delta$ above its ground state. When the projectile 
energy $E$ is above the threshold $\Delta$ then, apart from the elastic 
scattering the inelastic process is also possible. One may however describe the situation in terms of two channels, the elastic and the inelastic. The latter 
channel corresponds to the scattering of the projectile of energy $E>\Delta$ 
from the excited state of the system through a rectangular potential $V_I$ 
say of same width $L$, while the two channels are coupled to each other via a 
potential $V_c$ for simplicity, taken, to be that of same width $L$. With 
 $\psi(x)$ and $\phi(x)$ denoting the wave-functions in the elastic and the 
inelastic channels respectively, the system is governed by the coupled 
Schro\"dinger equations
$$
\left[-\frac{\hbar^2}{2m}\frac{d^2}{dx^2}+V_0\right]\psi(x)+V_c\phi(x)=E\psi(x)
\eqno(17a)
$$
$$
\left[-\frac{\hbar^2}{2m}\frac{d^2}{dx^2}+V_I\right]\phi(x)+V_c\psi(x)=
(E-\Delta)\phi(x)
\eqno(17b)
$$
The wave-functions for the elastic channel and the inelastic channel in the 
region $x<L$ is given as
$$
\psi(x)=e^{ikx}+a_R(k)e^{-ikx}\\
\eqno(18a)
$$
$$
\phi(x)=Re^{-ik^\prime x}	
\eqno(18b)
$$
 where $k^2=\frac{2mE}{\hbar^2}$ and $k^\prime 2=\frac{2m(E-\Delta)}{\hbar^2}$  
while $a_R(k)$ and $R$ are the respective reflection amplitudes in the elastic 
and inelastic channel. Similarly the wave-functions in the region $x>L$ is 
$$
\psi(x)=a_T(k)e^{ikx}\\
\eqno(19a)
$$
$$
\phi(x)=Te^{ik^\prime x}
\eqno(19b)
$$
where $a_T$ and $T$ are the respective transmission amplitudes in the elastic 
and inelastic channels. Note here that naturally there are two outgoing waves 
to the left and to the right due to the  additional inelastic channel as 
physically required.  The wave-functions in the two channels in the 
intermediate region are
$$
\psi(x)=(Be^{\alpha x}+Ce^{-\alpha x})sin(\theta/2)+(Fe^{\beta x}+Ge^{-\beta x})cos(\theta/2)
\eqno(20a)
$$
$$
\phi(x)=(Be^{\alpha x}+Ce^{-\alpha x})cos(\theta/2)-(Fe^{\beta x}+Ge^{-\beta x}
)sin(\theta/2)
\eqno(20b)
$$
where 
$$
\alpha^2=\frac{V_0+V_I-2E+\Delta}{2}+\sqrt{(V_0-V_I-\Delta)^2/4+V_c^2}
\eqno(21a)
$$
$$
\beta^2=\frac{V_0+V_I-2E+\Delta}{2}-\sqrt{(V_0-V_I-\Delta)^2/4+V_c^2}
\eqno(21b)
$$
$$
\theta=\frac{2V_c}{V_0-V_I-\Delta}
\eqno(21c)
$$
The phase difference between the incident and the transmitted waves in the 
elastic channel can be easily computed from $a_T$ evaluated by demanding the 
continuity of the wave-function and its derivative at the boundaries and hence 
the tunneling time in the elastic channel is expressed as a function of 
$\alpha$ and $\beta$. The two channels, being coupled by a coupling potential 
$V_c$, must admit evanescent wave-functions inside the potential barriers of 
both the channels for Hartman-Fletcher effect to be present in the elastic 
channel. For the wave-function to be evanescent inside, $\alpha$ and $\beta$ 
must be real. Now
$$
\alpha\beta=\sqrt{(V_I-E+\Delta)(V_0-E)-V_c^2}.
\eqno(22)
$$                    
For weak coupling (i.e., low $V_c$) and 'opaque condition' in the inelastic 
channel, i.e, $V_I>E-\Delta$, we have $\alpha\beta>0$ and real. If the opaque 
condition in both the channels are to be respected along with weak coupling, 
then both $\alpha$ and $\beta$ cannot simultaneously be purely imaginary. Hence both $\alpha$ and $\beta$ are real as a result of which evanescent wave 
occurs inside the barriers in the two-channels. However, if $V_I<E-\Delta$, 
then $\alpha\beta$ is imaginary and both or either of them must be imaginary 
resulting in the absence of evanescent wave inside the barrier in the elastic 
channel. In general, Hartman-Fletcher effect is expected if
$$
(V_I-E-\Delta)(V_0-E)>V_c^2
\eqno(23)
$$
Instead, if an attractive potential is considered in the inelastic channel, 
which is physically not forbidden, we have $V_I=-V_I^0$ (say), then
$$
\alpha\beta=\sqrt{(-V_I^0-E+\Delta)(V_0-E)-V_c^2}
\eqno(24)
$$
If the inelastic channel is to be kept open, i.e., $E>\Delta$, then 
$\alpha\beta$ is imaginary even for a weak coupling. So either or both of 
$\alpha,\beta$ are imaginary, as a result of which no evanescent wave, and 
hence the Hartman-Fletcher effect is expected. 

We have thus demonstrated that the independence of the tunneling time 
with respect to the width ($L$) of a barrier for a real potential with large 
$L$, claimed to be violated if inelastic channels are present, is a statement 
which needs to be qualified. We show that if instead of trying to capture 
inelasticity through the device of a complex potential, we treat the 
process through a multichannel formalism (more appropriate for the study of 
time evolution as compared to approaches using on-Hermitian Hamiltonians), then the outcome depends on some details of the channels and the strength of the 
coupling between them.

\vskip .5cm
\noindent $\ast${\footnotesize present address: Max-Plank-Institut f\"ur Physik komplexer Systeme,
                N\"othnitzer Stra$\beta$e 38, D-01187 Dresden, Germany.}

\begin{center}
{\bf References}
\end{center}
\vskip .3cm
\noindent [1]  L.Eisenbud, Ph.D. dissertation, Princeton University, 1948.\\

\noindent [2] E.P.Wigner, {\it Phys. Rev.}, {\bf 98}, 145 (1955).\\

\noindent [3] D.Bohm, {\it Quantum Theory}, Dover Publications.\\ 

\noindent [4] T.E.Hartman, {\it J. Appl. Phys}, {\bf 33} 3427 (1962).\\

\noindent [5] J.R.Fletcher, {\it J. Phys.} C {\bf 18}, L55 (1985).\\

\noindent [6] F. Delgado, G.Muga, and A.Rushhaupt, 
{\it Phys Rev A} {\bf 69}, 022106, 2004. \\ 

\noindent [7] F.Raciti and G.Salesi, {\it J.Phys. (France)}, 
{\bf 4}, 1783, 1994. \\ 

\noindent [8] G.Nimtz, {\it Proceedings of the Erice International Course:  
"Advances in Quantum Mechanics"}, 1994; and G.Nimtz, H.Spieker and 
H.M.Brodowsky, {\it "Tunneling with Dissipation"-preprint}.

\end{document}